\documentclass[aps,prd,twocolumn,tightenlines,superscriptaddress,showpacs,byrevtex]{revtex4}

\usepackage{graphicx} 
\usepackage{dcolumn}  

\graphicspath{{ps}}

\newcommand{\BF}{\ensuremath{{\cal{B}}}}
\newcommand{\Am}{\ensuremath{{\cal{A}}}}
\newcommand{\UFS}{\ensuremath{\Upsilon(4S)}}
\newcommand{\bbbar}{\ensuremath{B\bar{B}}}
\newcommand{\qqbar}{\ensuremath{q\bar{q}}}

\newcommand{\de}{\ensuremath{\Delta E}}
\newcommand{\mb}{\ensuremath{M_{\rm bc}}}

\newcommand{\mass}{MeV/$c^2$}

\newcommand{\bckpp}{\ensuremath{B^+\to K^+\pi^+\pi^-}}
\newcommand{\bckkk}{\ensuremath{B^+\to K^+K^+K^-}}
\newcommand{\bskpp}{\ensuremath{B^0\to K^0_S\pi^+\pi^-}}
\newcommand{\bnkpp}{\ensuremath{B^0\to K^0\pi^+\pi^-}}

\newcommand{\kspp}{\ensuremath{K^0_S\pi^+\pi^-}}
\newcommand{\knpp}{\ensuremath{K^0\pi^+\pi^-}}
\newcommand{\kcpp}{\ensuremath{K^+\pi^+\pi^-}}

\newcommand{\pipi}{\ensuremath{\pi^+\pi^-}}

\newcommand{\kspi}{\ensuremath{K^0_S\pi^{\pm}}}

\newcommand{\ks}{\ensuremath{K^0_S}}

\newcommand{\chic}{\ensuremath{\chi_{c0}}}

\newcommand{\sfs}{\ensuremath{s_{12}}}
\newcommand{\sft}{\ensuremath{s_{13}}}
\newcommand{\sst}{\ensuremath{s_{23}}}

\newcommand{\lumi}{\ensuremath{357~{\rm fb}^{-1}}}

\def\nim#1#2#3{{Nucl.\ Instr.\ and Meth.} {\bf #1}, #3 (#2)}
\def\nima#1#2#3{{Nucl.\ Instr.\ and Meth.} A {\bf #1}, #3 (#2)}

\def\jpg#1#2#3{{ J.\ Phys.}        G {\bf #1}, #3 (#2)}
\def\npb#1#2#3{{ Nucl.\ Phys.}     B {\bf #1}, #3 (#2)}
\def\plb#1#2#3{{ Phys.\ Lett.}     B {\bf #1}, #3 (#2)}

\def\prd#1#2#3{{ Phys.\ Rev.}      D {\bf #1}, #3 (#2)}

\def\prl#1#2#3{{ Phys.\ Rev.\ Lett.} {\bf #1}, #3 (#2)}

\def\zpc#1#2#3{{ Zeit.\ Phys.}     C {\bf #1}, #3 (#2)}
\begin{document}

\title{ Dalitz Analysis of Three-body Charmless $\bnkpp$ Decay}


\affiliation{Budker Institute of Nuclear Physics, Novosibirsk}
\affiliation{Chiba University, Chiba}
\affiliation{Chonnam National University, Kwangju}
\affiliation{University of Cincinnati, Cincinnati, Ohio 45221}
\affiliation{Department of Physics, Fu Jen Catholic University, Taipei}
\affiliation{The Graduate University for Advanced Studies, Hayama, Japan} 
\affiliation{University of Hawaii, Honolulu, Hawaii 96822}
\affiliation{High Energy Accelerator Research Organization (KEK), Tsukuba}
\affiliation{Hiroshima Institute of Technology, Hiroshima}
\affiliation{Institute of High Energy Physics, Chinese Academy of Sciences, Beijing}
\affiliation{Institute of High Energy Physics, Vienna}
\affiliation{Institute of High Energy Physics, Protvino}
\affiliation{Institute for Theoretical and Experimental Physics, Moscow}
\affiliation{J. Stefan Institute, Ljubljana}
\affiliation{Kanagawa University, Yokohama}
\affiliation{Korea University, Seoul}
\affiliation{Kyungpook National University, Taegu}
\affiliation{Swiss Federal Institute of Technology of Lausanne, EPFL, Lausanne}
\affiliation{University of Ljubljana, Ljubljana}
\affiliation{University of Maribor, Maribor}
\affiliation{University of Melbourne, Victoria}
\affiliation{Nagoya University, Nagoya}
\affiliation{Nara Women's University, Nara}
\affiliation{National Central University, Chung-li}
\affiliation{National United University, Miao Li}
\affiliation{Department of Physics, National Taiwan University, Taipei}
\affiliation{H. Niewodniczanski Institute of Nuclear Physics, Krakow}
\affiliation{Nippon Dental University, Niigata}
\affiliation{Niigata University, Niigata}
\affiliation{University of Nova Gorica, Nova Gorica}
\affiliation{Osaka City University, Osaka}
\affiliation{Osaka University, Osaka}
\affiliation{Panjab University, Chandigarh}
\affiliation{Peking University, Beijing}
\affiliation{Princeton University, Princeton, New Jersey 08544}
\affiliation{RIKEN BNL Research Center, Upton, New York 11973}
\affiliation{University of Science and Technology of China, Hefei}
\affiliation{Seoul National University, Seoul}
\affiliation{Shinshu University, Nagano}
\affiliation{Sungkyunkwan University, Suwon}
\affiliation{University of Sydney, Sydney NSW}
\affiliation{Tata Institute of Fundamental Research, Bombay}
\affiliation{Toho University, Funabashi}
\affiliation{Tohoku Gakuin University, Tagajo}
\affiliation{Tohoku University, Sendai}
\affiliation{Department of Physics, University of Tokyo, Tokyo}
\affiliation{Tokyo Institute of Technology, Tokyo}
\affiliation{Tokyo Metropolitan University, Tokyo}
\affiliation{Tokyo University of Agriculture and Technology, Tokyo}
\affiliation{Virginia Polytechnic Institute and State University, Blacksburg, Virginia 24061}
\affiliation{Yonsei University, Seoul}
  \author{A.~Garmash}\affiliation{Princeton University, Princeton, New Jersey 08544} 
  \author{K.~Abe}\affiliation{High Energy Accelerator Research Organization (KEK), Tsukuba} 
  \author{I.~Adachi}\affiliation{High Energy Accelerator Research Organization (KEK), Tsukuba} 
  \author{H.~Aihara}\affiliation{Department of Physics, University of Tokyo, Tokyo} 
  \author{D.~Anipko}\affiliation{Budker Institute of Nuclear Physics, Novosibirsk} 
  \author{K.~Arinstein}\affiliation{Budker Institute of Nuclear Physics, Novosibirsk} 
  \author{V.~Aulchenko}\affiliation{Budker Institute of Nuclear Physics, Novosibirsk} 
  \author{T.~Aushev}\affiliation{Swiss Federal Institute of Technology of Lausanne, EPFL, Lausanne}\affiliation{Institute for Theoretical and Experimental Physics, Moscow} 
  \author{S.~Bahinipati}\affiliation{University of Cincinnati, Cincinnati, Ohio 45221} 
  \author{A.~M.~Bakich}\affiliation{University of Sydney, Sydney NSW} 
% ...
\collaboration{The Belle Collaboration}

\noaffiliation

\begin{abstract}
     We report results of a Dalitz plot analysis of the three-body charmless
$\bnkpp$ decay. The analysis is performed with a data sample
that contains 388 million $B\bar{B}$ pairs collected near the $\Upsilon(4S)$
resonance with the Belle detector at the KEKB asymmetric energy $e^+e^-$
collider. Measurements of branching fractions for the quasi-two-body decays
$B^0\to\rho(770)^0K^0$, $B^0\to f_0(980)K^0$, $B^0\to K^*(892)^+\pi^-$,
$B^0\to K^*(1430)^+\pi^-$, and upper limits on several other quasi-two-body
decay modes are reported.
\end{abstract}

\pacs{13.20.He, 13.25.Hw, 13.30.Eg, 14.40.Nd}  

\maketitle

{\renewcommand{\thefootnote}{\fnsymbol{footnote}}}
\setcounter{footnote}{0}

\normalsize

\section{Introduction}

Decays of $B$ mesons to three-body charmless hadronic final states have
attracted considerable attention in recent years. An amplitude analysis for
a number of three-body final states has been performed
(for example, $K^+K^+K^-$, $K^+\pi^+\pi^-$, $K^+\pi^-\pi^0$),
where branching fractions for many quasi-two-body intermediate states have
been measured for the first time or with a significantly improved accuracy.

In addition to providing a rich laboratory for studying $B$ meson decay
dynamics, three-body charmless final states open new possibilities for $CP$
violation studies. Several new ideas utilizing three-body final states have
been proposed~\cite{hhh-theory}. Experimentally, studies of $CP$ violation
have been done with most of the final states mentioned above, yielding some
interesting results. For example, the first evidence for direct $CP$ violation
in charged $B$ meson decays to the $\rho(770)^0K^\pm$ final states has been
recently found through the amplitude analysis of the three-body
$B^\pm\to K^\pm\pi^\pm\pi^\mp$ decay~\cite{belle-dcpv}.
Time-dependent $CP$ violation was measured in
$B^0\to K^+K^-K^0$~\cite{b2s-belle,kskckc-babar} and
$B^0\to\ks\ks\ks$~\cite{ksksks-belle,ksksks-babar} three-body decays, which
occur dominantly via the $b\to s$ penguin transition. Measurements of
$\sin2\phi_1$ in $b\to s$ penguin-dominated decays provide an important test
of the Standard Model. The quasi-two-body $B^0\to f_0(980)\ks$ channel that
contributes to the three-body $\kspp$ final state is also expected to be
dominated by the $b\to s$ penguin transition and thus has been used for the
measurement of $\sin2\phi_1$~\cite{b2s-belle,f980-babar, note-1}. However,
since the $f_0(980)$ has a significant natural width, nearby resonant states
(for example the $\rho(770)^0$ is particularly important as the combined $CP$
parity of the $B^0\to \rho(770)\ks$ is opposite to that of the
$B^0\to f_0(980)\ks$) might contribute to the $f_0(980)$ mass region and an
accurate estimation of these contributions is required for a correct
interpretation of the results. This can only be done via an amplitude (Dalitz)
analysis of the three-body $\bnkpp$ decay.

In this paper we report first results of a Dalitz plot analysis of the
three-body charmless $\bnkpp$ decay. The analysis is based on a $\lumi$ data
sample containing $388\times10^{-6}$ $B\bar{B}$ pairs, collected with the Belle
detector operating at the KEKB asymmetric-energy $e^+e^-$ collider~\cite{KEKB}
with a center-of-mass (c.m.) energy at the $\Upsilon(4S)$ resonance. For the
study of the $e^+e^-\to q\bar{q}$ continuum background, we use a data sample
(that amounts to about 10\% of the on-resonance sample) taken 60~MeV below
the $\Upsilon(4S)$ resonance.

\section{The Belle detector}

The Belle detector~\cite{Belle} is a large-solid-angle magnetic spectrometer
based on a 1.5~T superconducting solenoid magnet. Charged particle tracking is
provided by a silicon vertex detector and a 50-layer central
drift chamber (CDC) that surround the interaction point. 
Charged hadron identification is provided by $dE/dx$ measurements in the CDC,
an array of 1188 aerogel \v{C}erenkov counters (ACC), and a barrel-like array
of 128 time-of-flight scintillation counters (TOF); information from the three
subdetectors is combined to form a single likelihood ratio for each pair of
hadron species that is then used
for pion, kaon and proton discrimination. Electromagnetic showering
particles are detected in an array of 8736 CsI(Tl) crystals (ECL) that covers
the same solid angle as the charged particle tracking system.
Electron identification is based on a combination of $dE/dx$
measurements in the CDC, the response of the ACC, and the position, shape and
total energy deposition of the shower detected in the ECL.
The electron identification efficiency is greater than 92\% for tracks with
$p_{\rm lab}>1.0$~GeV/$c$ and the hadron misidentification probability is below
0.3\%. The magnetic field is returned via an iron yoke that is instrumented to
detect muons and $K^0_L$ mesons. We use a GEANT-based Monte Carlo (MC)
simulation to model the response of the detector and determine its
acceptance~\cite{GEANT}.


\section{Event Reconstruction}

Candidate charged pions from $B$ meson decay are required to be consistent with
having originated from the interaction point and to have momenta transverse to
the beam greater than 0.1~GeV/$c$. To reduce the combinatorial background, we
impose a requirement on the particle identification variable that has 93\%
efficiency and about 15\% fake rate from misidentified kaons. Tracks that are
positively identified as electrons or protons are excluded. We fit these
candidate pions to the common vertex to determine the $B$ meson decay vertex.
Neutral kaons are reconstructed via the decay $K^0\to\pi^+\pi^-$. The invariant
mass of the two oppositely charged tracks is required to be within 12~MeV/$c^2$
of the nominal $\ks$ mass. The direction of flight of the $\ks$ candidate is
required to be consistent with the direction of its vertex displacement with
respect to the $B$ decay vertex. 

$B$ candidates are identified using two kinematic variables: 
the beam-constrained mass
$\mb=\frac{1}{c^2}\sqrt{E^{*2}_{\rm beam}-c^2|\sum_i\mathbf{p}_i|^2},$
and the energy difference
$\de=(\sum_i\sqrt{c^2|\mathbf{p}_i|^2 + c^4m_i^2} ) - E^*_{\rm beam},$
where the summation is over all particles from a $B$ candidate; ${\bf p}_i$
and $m_i$ are their c.m.\ three-momenta and masses, respectively;
$E^*_{\rm beam}$ is the beam energy in the c.m.\, frame. The signal $\mb$
resolution is mainly determined by the beam energy spread and amounts to
2.9~MeV/$c^2$. The signal $\de$ shape is fit to a sum of two Gaussian
functions (core and tail) with a common mean. 


\begin{figure}[!t]
 \includegraphics[width=0.49\textwidth]{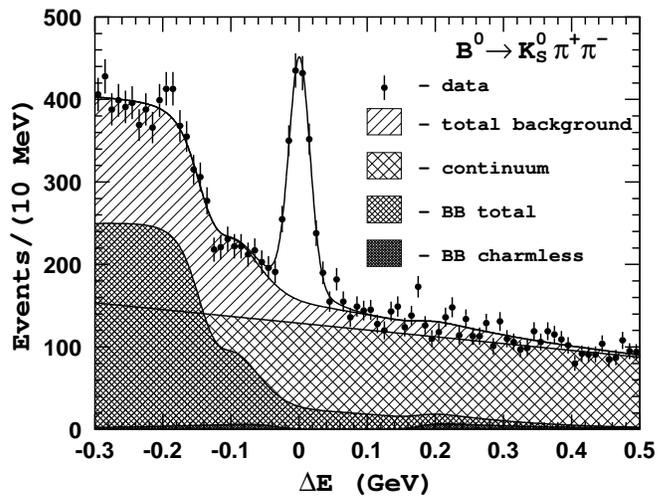}
 \caption{$\de$ distribution for the $\bnkpp$ candidate events with
          $|\mb-M_B|<7.5$~MeV/$c^2$. Points with error bars are data; the
          upper curve is the fit result; the hatched histograms are various
          background components.}
 \label{fig:dE-Mbc}
\end{figure}

The dominant background is due to $e^+e^-\to~\qqbar$ ($q = u, d, s$ and $c$
quarks) continuum events. We suppress this background using variables that
characterize the event topology. Since the two $B$ mesons produced from an
$\UFS$ decay are nearly at rest in the c.m.\ frame, their decay products are
uncorrelated and the event tends to be spherical. In contrast, hadrons from
continuum $\qqbar$ events tend to exhibit a two-jet structure. We use
$\theta_{\rm thr}$, which is the angle between the thrust axis of the $B$
candidate and that of the rest of the event, to discriminate between the two
cases. The distribution of $|\cos\theta_{\rm thr}|$ is strongly peaked near
$|\cos\theta_{\rm thr}|=1.0$ for $\qqbar$ events and is nearly flat for
$\bbbar$ events. We require $|\cos\theta_{\rm thr}|<0.80$ eliminating about
83\% of the continuum background while retaining 79\% of the signal events. For
further suppression of the continuum background, we use a Fisher discriminant
formed from 11 variables: nine variables that characterize the angular
distribution of the momentum flow in the event with respect to the $B$
candidate thrust axis, the angle of the $B$ candidate thrust axis with respect
to the beam axis, and the angle between the $B$ candidate momentum and the
beam axis.  Use of such a Fisher discriminant rejects about 89\% of the
remaining continuum background with 53\% efficiency for the signal. A more
detailed description of the background suppression technique can be found in
Ref.~\cite{belle-khh2} and references therein.


\begin{figure*}[!th]
 \includegraphics[width=0.49\textwidth]{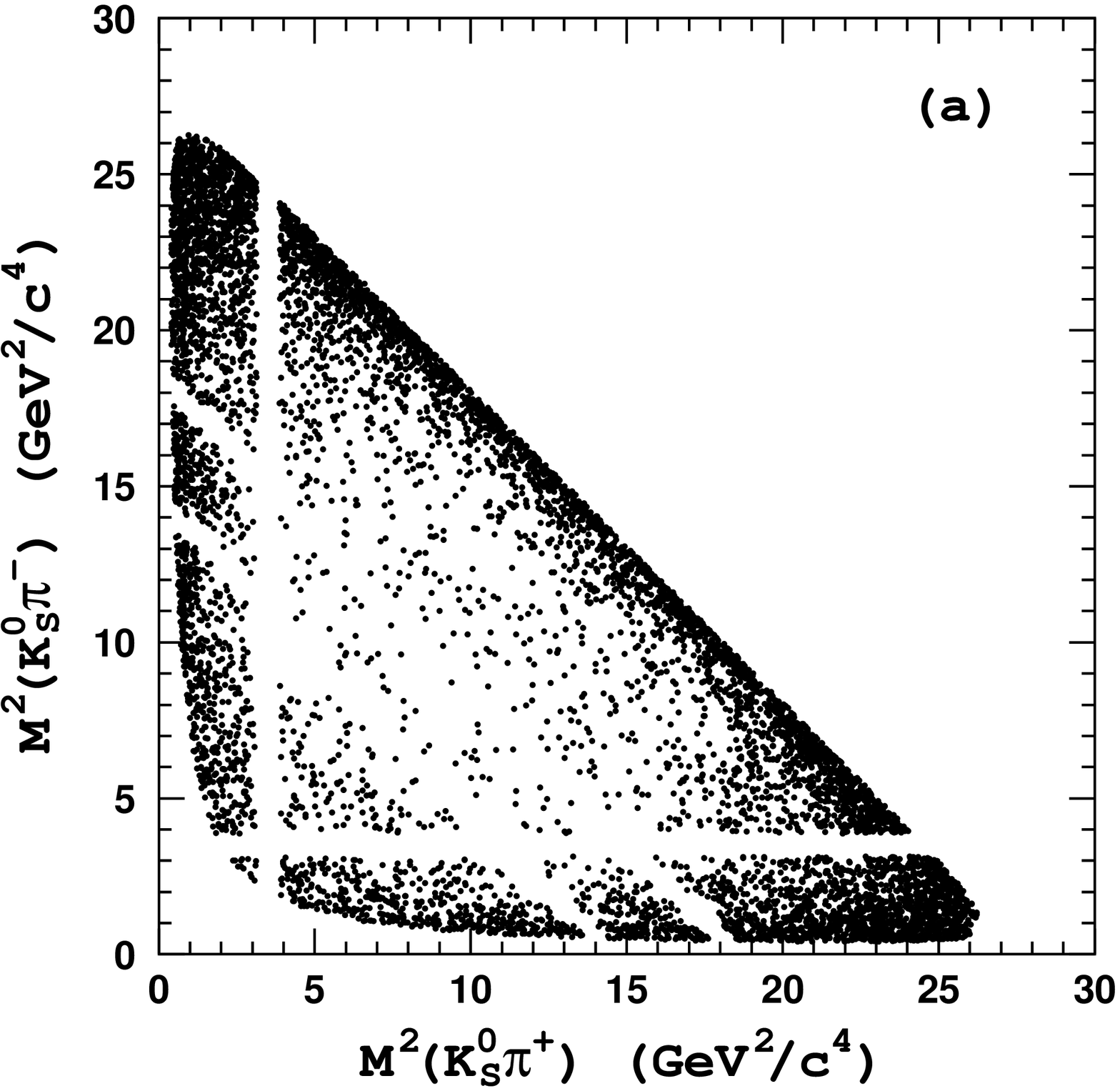} \hfill
 \includegraphics[width=0.49\textwidth]{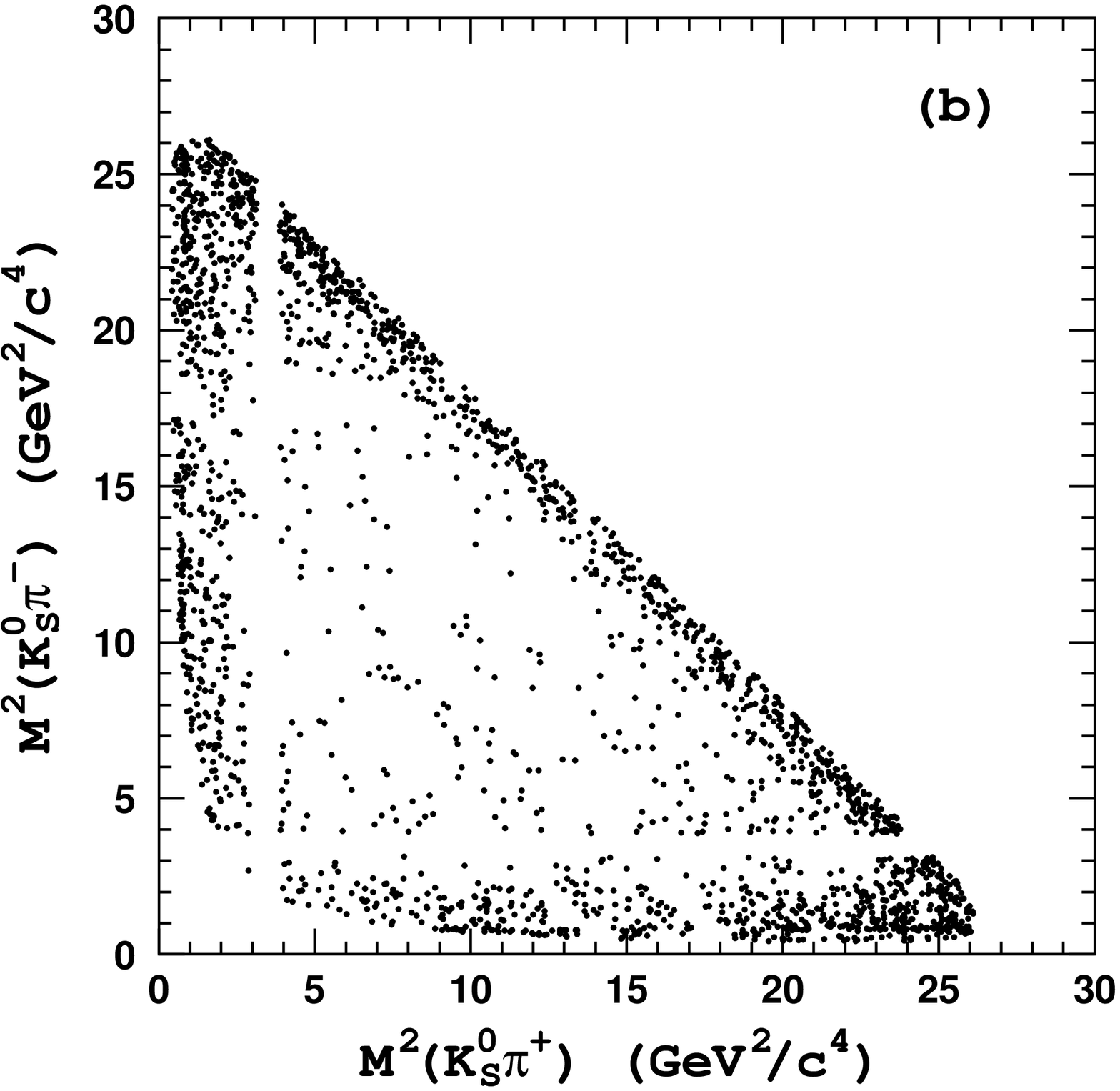}
 \caption{Dalitz plots for events in the (a) $\mb$$-$$\de$ sidebands and in
          the (b) $B$ signal region.}
 \label{fig:DP}
\end{figure*}

From MC study we find that the backgrounds originating from other $B$ meson
decays that peak in the signal region are due to $B^0\to D^-[\ks\pi^-]\pi^+$
as well as $B^0\to J/\psi[\mu^+\mu^-]\ks$ and $B^0\to\psi(2S)[\mu^+\mu^-]\ks$
decays with muons misidentified as pions. We veto these backgrounds by
requiring $|M(\ks\pi^-)-M_D|>100$~MeV/$c^2$,
$|M(\pipi)_{\mu\mu}-M_{J/\psi}|>70$~MeV/$c^2$ and
$|M(\pipi)_{\mu\mu}-M_{\psi(2S)}|>50$~MeV/$c^2$, with a muon mass assignment
used here for the pion candidates. To suppress the background due to $K/\pi$
misidentification, we exclude candidates that are consistent with the
$D^-\to\ks K^-$ hypothesis within 15~MeV/$c^2$ ($\sim 2.5\sigma$), regardless
of the particle identification information. There is also a large background
from the $B\to D[K\pi\pi]\pi$ channel and from the $B\to D^{(*)}\pi$ channel
with a subsequent semileptonic $D\to K\mu\nu_\mu$ decay. However, these modes
do not peak in the signal region and contribute mainly to the $\de<0$ region.
The most significant backgrounds from charmless $B$ decays originate from
$B^0\to\eta'[\pi^+\pi^-\gamma]\ks$ and $B^\pm\to\ks\pi^\pm$ decays. In the
latter case an additional soft pion is randomly picked up to form a $\kspp$
combination. We determine the $\de$ shape for these backgrounds from MC
simulation and take them into account when fitting the data.

The $\de$ distribution for $\kspp$ combinations that pass all the selection
requirements is shown in Fig.~\ref{fig:dE-Mbc}, where a clear peak in the
signal region is observed. In the fit to the $\de$ distribution we fix the
shape of the $\bbbar$ background component from MC and let the normalization
float. The shape of the $\qqbar$ background is parametrized by a linear
function with slope and normalization as free fit parameters. For the signal
component the width ($\sigma$) and the relative fraction of the tail Gaussian
function are fixed at $30$~MeV and $0.19$, respectively, as determined from
signal MC simulation. The common mean of the two Gaussian functions and the
width of the core Gaussian are allowed to float and found to be $0.7\pm0.6$~MeV
and $15.3\pm0.9$~MeV, respectively. The fit yields $1229\pm62$ signal $\bskpp$
events.


\section{Amplitude Analysis}

The amplitude analysis of the three-body $\bnkpp$ signal is performed by means
of an unbinned maximum likelihood fit. In general, we follow the procedure we
used for the analysis of the decay $\bckpp$ described in detail in
Ref.~\cite{belle-khh3}. For the analysis we select events in the $B$ signal
region defined as an ellipse around the $\mb$ and $\de$ signal mean values:
\[
 \left[\frac{\mb-M_B}{7.5~{\rm MeV}/c^2}\right]^2+
 \left[\frac{\de}{40~{\rm MeV}}\right]^2<1. \nonumber
\]
To determine the distribution of background events over the phase space
(Dalitz plot) we use events in the $\mb-\de$ sidebands defined as
\begin{eqnarray}
0.05{\rm ~GeV}/c^2<|M(K\pi\pi)-M_B|<0.10{\rm ~GeV}/c^2; \nonumber \\
P(K\pi\pi)<0.48{\rm ~GeV}/c~~~~~~~~~~~~~~~~~ \nonumber
\end{eqnarray}
and
\begin{eqnarray}
|M(K\pi\pi)-M_B|<0.10{\rm ~GeV}/c^2;~~~ \nonumber \\
0.48{\rm ~GeV}/c<P(K\pi\pi)<0.65{\rm ~GeV}/c, \nonumber
\end{eqnarray}
where $M(K\pi\pi)$ and $P(K\pi\pi)$ are the three-particle invariant mass and
three-particle momentum in the c.m.~frame. The total number of events in the
signal (sideband) region is 2207 (8159). The relative fraction of signal
events in the signal region is $0.521\pm0.025$.

\subsection{Fit to Sideband Events}


\begin{figure}[!t]
  \centering
  \includegraphics[width=0.47\textwidth]{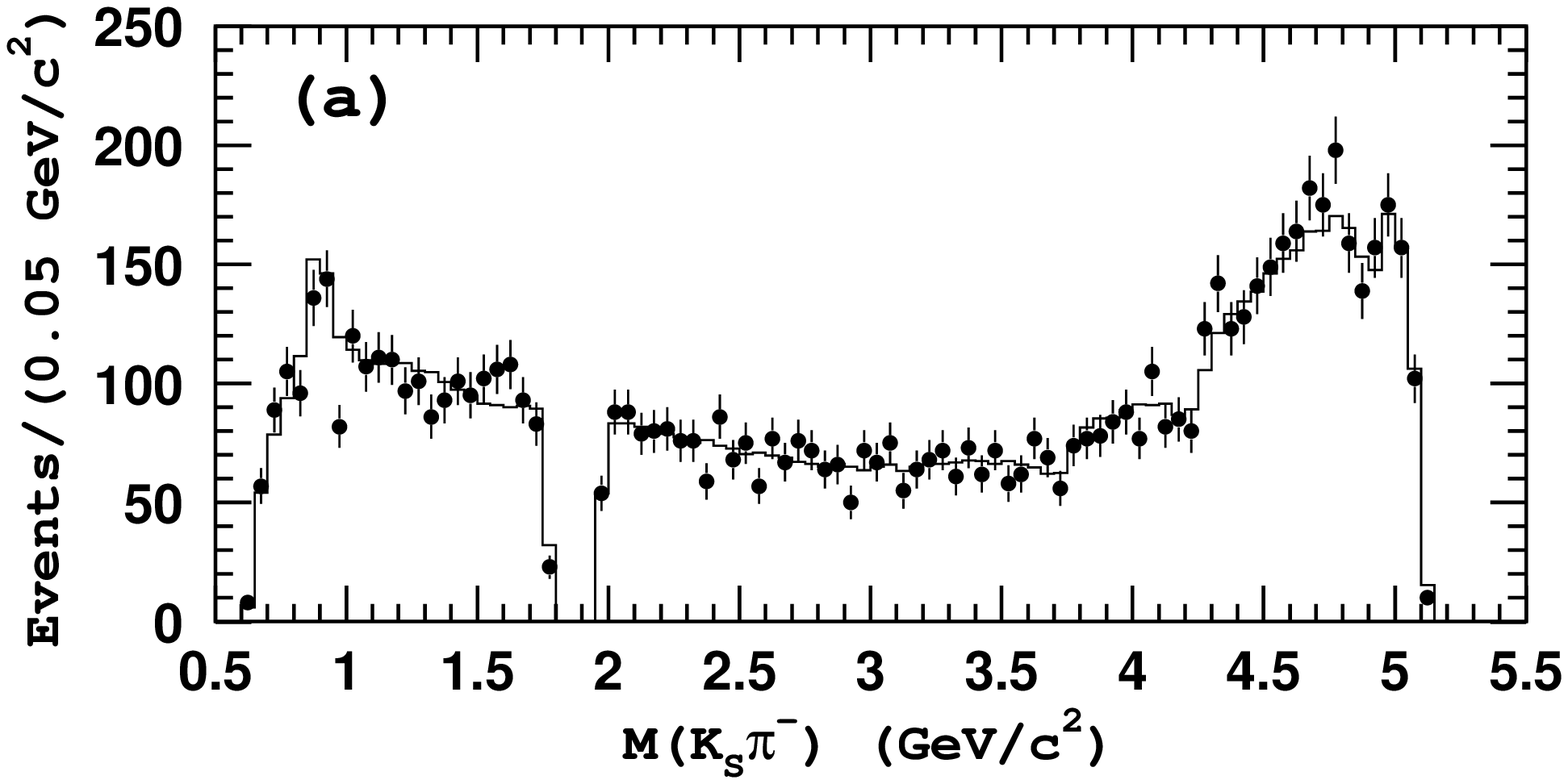}
  \includegraphics[width=0.47\textwidth]{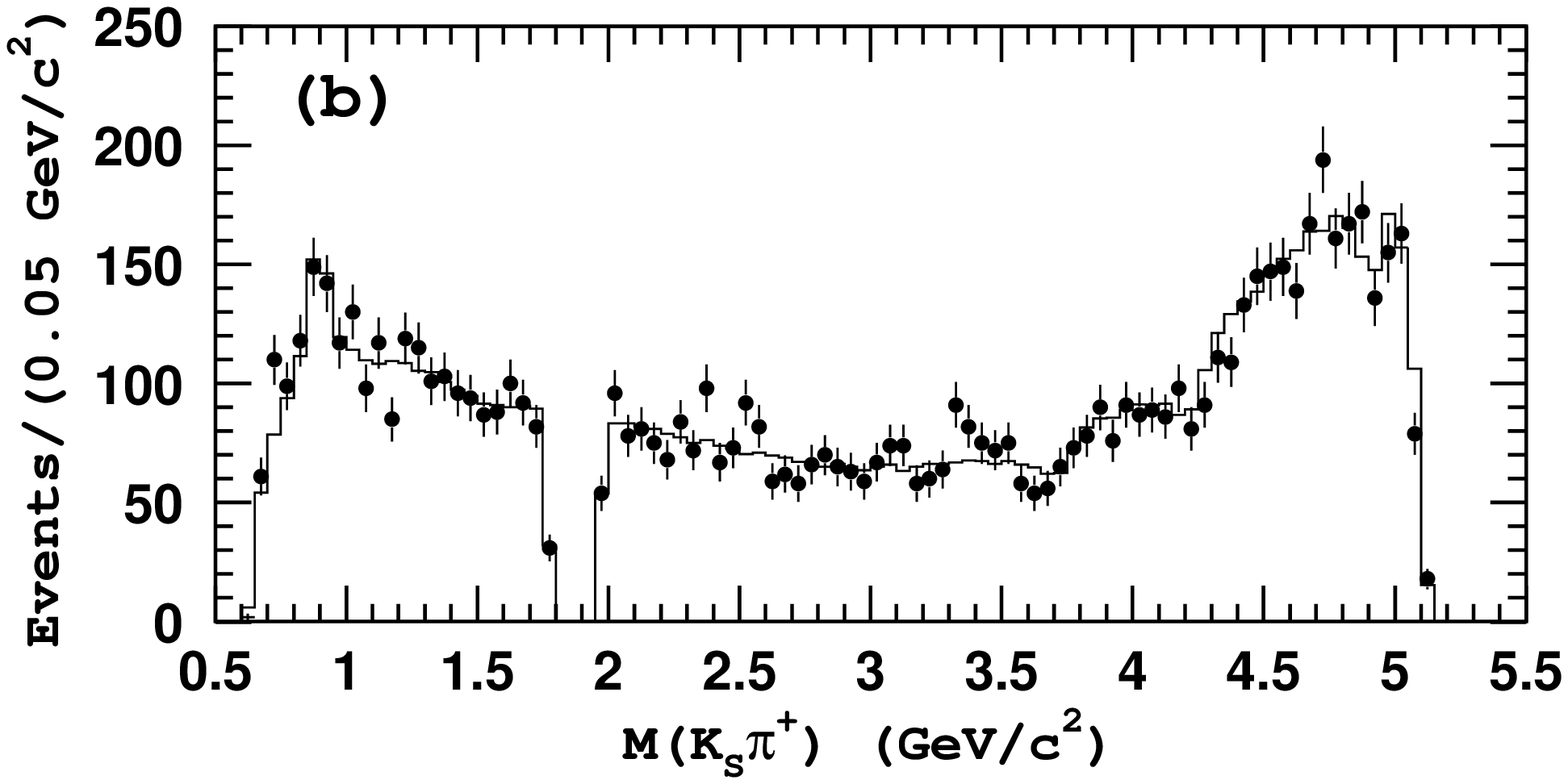}
  \includegraphics[width=0.47\textwidth]{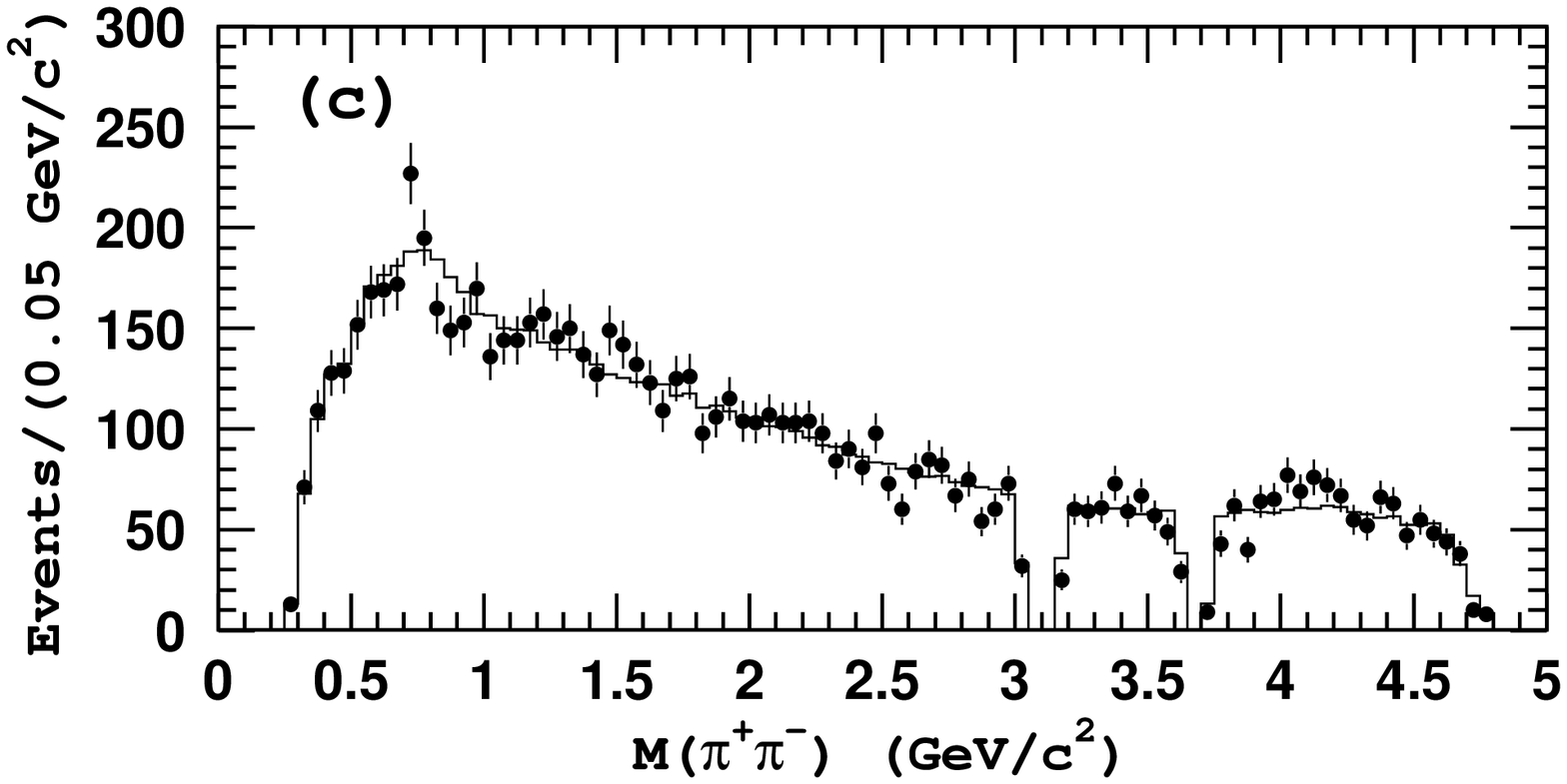}
  \caption{Results of the best fit to events in the $\de-\mb$ 
           sidebands. Points with error bars are data; histograms
           are fit results.}
\label{fig:khh_back}
\end{figure}

The Dalitz plot for events in the $\mb-\de$ sideband region is shown in
Fig.~\ref{fig:DP}(a) where visible gaps are due to vetoes applied on invariant
masses of two-particle combinations. We use the following empirical
parametrization to describe the distribution of background events over the
Dalitz plot:
\begin{eqnarray}
&&B(\ks\pi^\pm\pi^\mp)  = \alpha_1(e^{-\beta_1s_{12}}+e^{-\beta_1s_{13}})
         ~+~ \alpha_2e^{-\beta_2s_{23}} \nonumber \\
      &&~~+~ \alpha_3(e^{-\beta_3s_{12}-\beta_4s_{23}}
         ~+~          e^{-\beta_3s_{13}-\beta_4s_{23}}) \nonumber \\
      &&~~+~ \alpha_4e^{-\beta_5(s_{12}+s_{13})} \nonumber \\
      &&~~+~ \gamma_1(|BW_1(K^*(892)^-)|^2+|BW_1(K^*(892)^+)|^2) \nonumber \\
      &&~~+~ \gamma_2 |BW(\rho(770)^0)|^2,
\label{eq:kpp_back}
\end{eqnarray}
where $s_{12} \equiv M^2(\ks\pi^-)$,
$s_{13} \equiv M^2(\ks\pi^+)$, $s_{23}\equiv M^2(\pipi)$ and $\alpha_i$
($\alpha_1\equiv 1.0$), $\beta_i$ and $\gamma_i$ are fit parameters; $BW$
is a Breit-Wigner function. The first two terms in Eq.~(\ref{eq:kpp_back}) are
introduced to describe the excess of background events in the two-particle low
invariant mass regions (borders of the Dalitz plot). This enhancement
originates mainly from $e^+e^-\to\qqbar$ continuum events; due to the jet-like
structure of this background, all three particles in a three-body combination
have almost collinear momenta. Hence, the invariant mass of at least one pair
of particles is in the low mass region. In addition, it is often the case that
two high momentum particles are combined with a low momentum particle to form
a $B$ candidate. In this case there are two pairs with low invariant masses
and one pair with high invariant mass, resulting in even  stronger enhancement
of the background in the corners of the Dalitz plot. This is taken into account
by terms proportional to $\alpha_3$ and $\alpha_4$ in Eq.~(\ref{eq:kpp_back}).
To account for a possible contribution from real $K^*(892)^\pm$ and
$\rho(770)^0$ mesons, we introduce two more terms in Eq.~(\ref{eq:kpp_back}),
that are (non-interfering) squared Breit-Wigner amplitudes, with masses and
widths fixed at world average values~\cite{PDG}. The two-particle invariant
mass projections for the sideband data and the fit results are shown in
Fig.~\ref{fig:khh_back}.

\subsection{Fit to Signal Events}

\begin{figure}[t]
  \centering
  \includegraphics[width=0.48\textwidth]{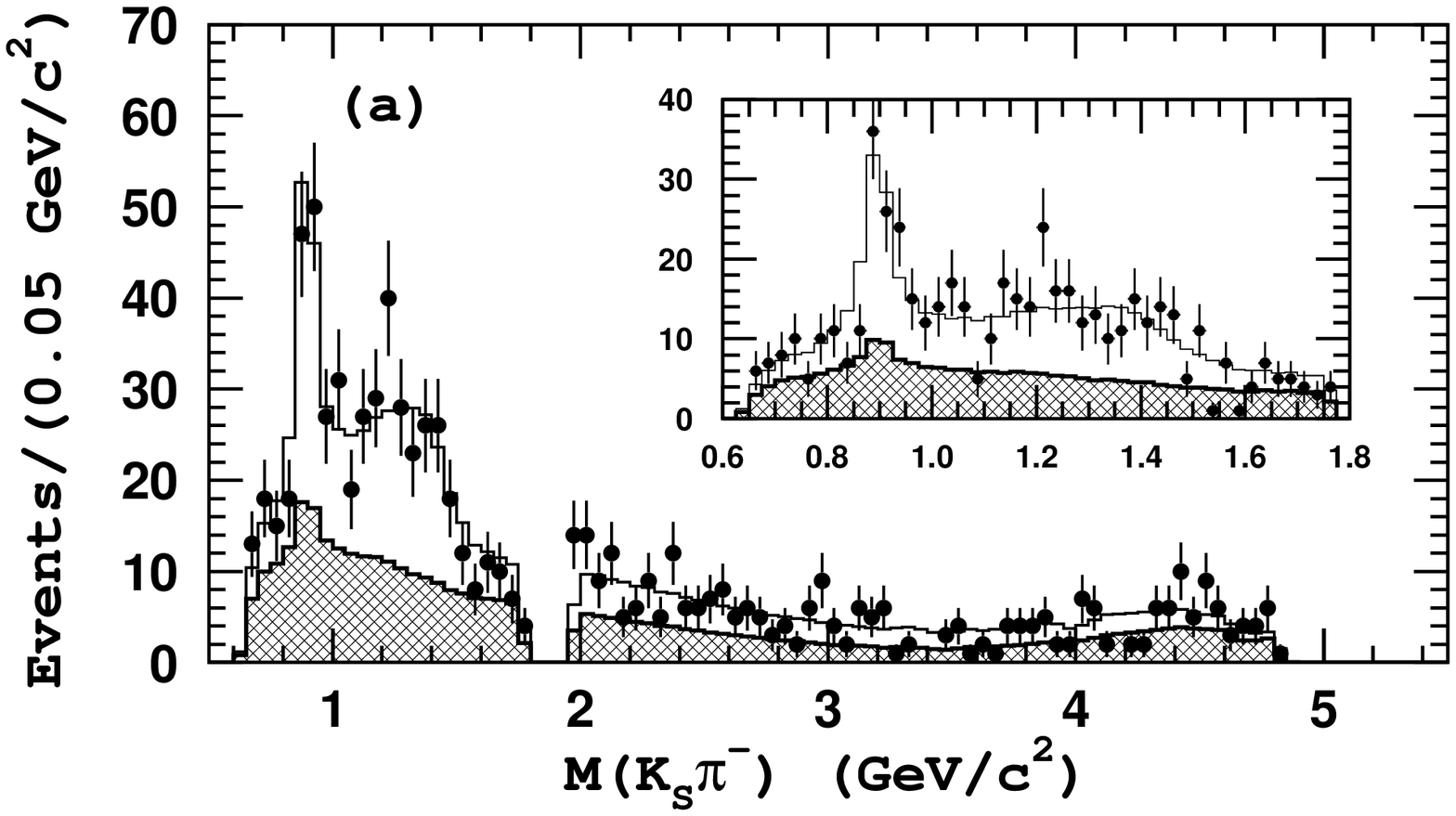}
  \includegraphics[width=0.48\textwidth]{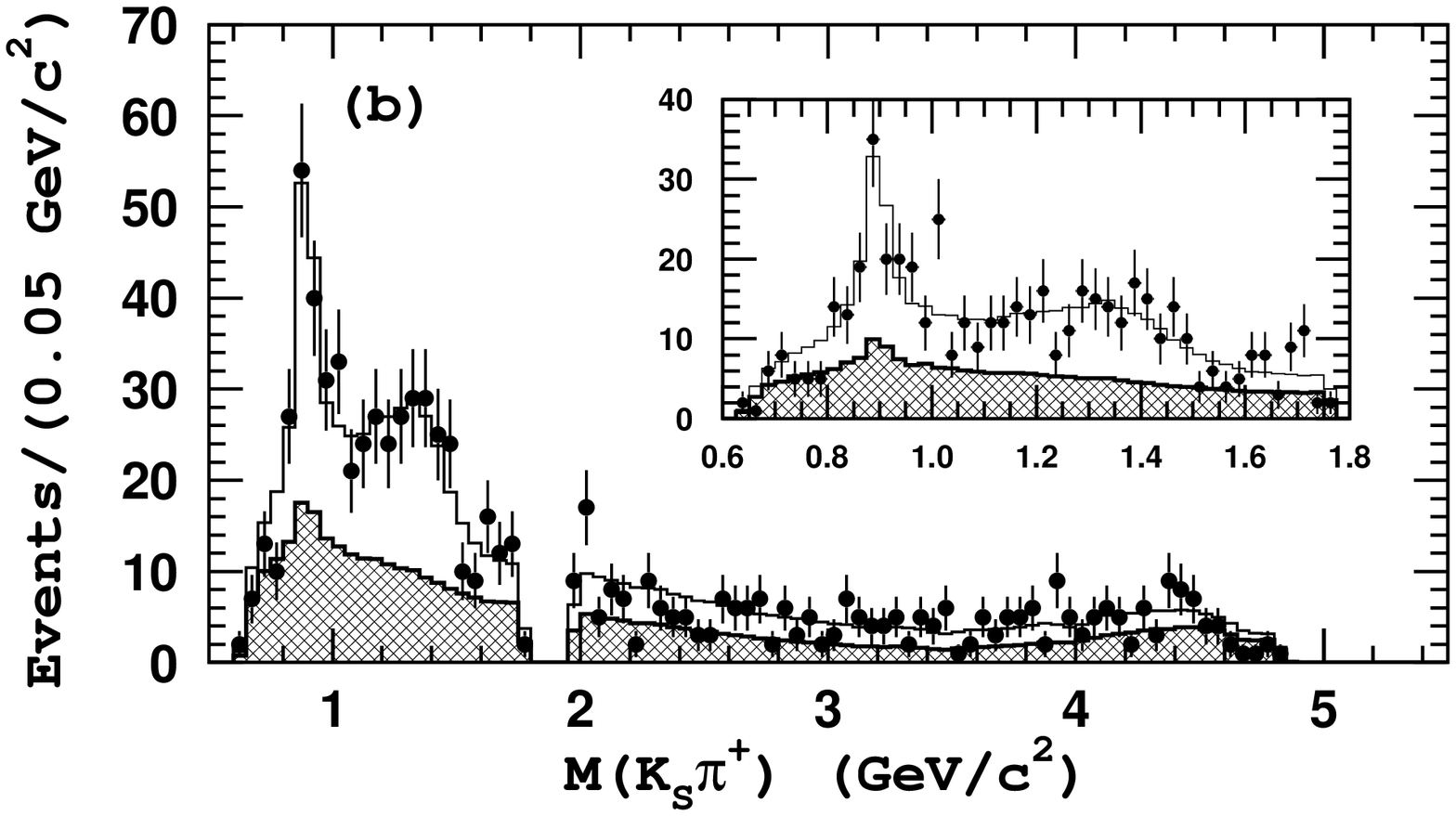}
  \includegraphics[width=0.48\textwidth]{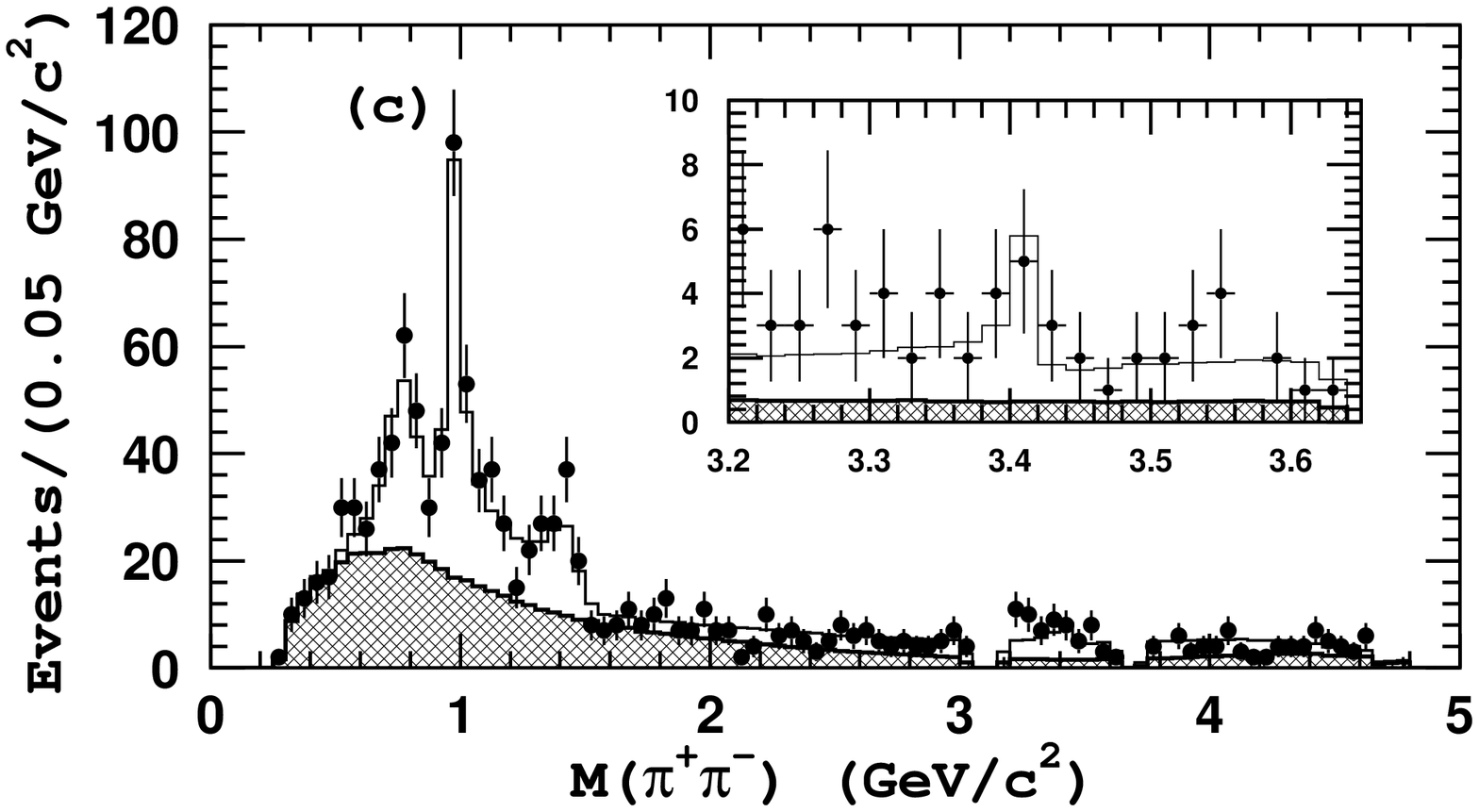}
  \caption{Results of the fit to $\kspp$ events in the signal region.
           Points with error bars are data, the open
           histograms are the fit result and hatched histograms are the
           background components. Insets in (a) and (b) show the
           $K^*(892)-K_0^*(1430)$ mass region in 20~\mass~ bins; inset in (c)
           shows the $\chic$ mass region in 25~\mass~ bins. When plotting a
           two-particle mass projection we require the invariant mass of the
           other two two-particle combinations to be greater than
           $1.5$~GeV/$c^2$.}
\label{fig:kpp-sig-fit}
\medskip
\end{figure}

The Dalitz plot for events in the signal region is shown in
Fig.~\ref{fig:DP}(b);
Figure~\ref{fig:kpp-sig-fit} shows the two-particle invariant mass
distributions. In an attempt to describe all the features of the $\kspi$ and
$\pipi$ mass spectra visible in Fig.~\ref{fig:kpp-sig-fit}, we use a matrix
element similar to that constructed in the analysis of the $\bckpp$
decay~\cite{belle-khh3}:
\begin{eqnarray}
{\cal{M}}(\knpp)
&=& a_{K^*}e^{i\delta_{K^*}}\Am_1(\pi^+K^0\pi^-|K^*(892)^+)\nonumber\\
&+& a_{K^*_0}e^{i\delta_{K^*_0}}\Am_0(\pi^+K^0\pi^-|K^*_0(1430)^+)\nonumber\\
&+& a_{\rho}e^{i\delta_{\rho}}\Am_1(K^0\pipi|\rho(770)^0) \nonumber\\
&+& a_{f_0}e^{i\delta_{f_0}}\Am_{\rm Flatte}(K^0\pipi|f_0(980)) \nonumber \\
&+& a_{f_X}e^{i\delta_{f_X}}\Am_0(K^0\pipi|f_X(1300)) \nonumber\\
&+& a_{\chic}e^{i\delta_{\chic}}\Am_0(K^0\pipi|\chic) \nonumber \\
&+& \Am_{\rm nr}(\knpp),
\label{eq:kpp-model}
\end{eqnarray}
where relative amplitudes $a_i$ and phases $\delta_i$ are fit parameters.
Each quasi-two-body amplitude $\Am_J $ is parametrized as
\begin{equation}
\Am_J = F_B F^{(J)}_R BW_J T_J,
\end{equation}
where $J$ is the spin of an intermediate resonant state; $BW_J$ is the
Breit-Wigner function; $F_B$ is the $B$ meson decay form factor parametrized
in a single-pole approximation~\cite{ffactor}; $F^{(J)}_R$ is the
Blatt-Weisskopf form factor~\cite{blatt-weisskopf} for the intermediate
resonance decay; and $T_J$ is the function that describes angular correlations
between final state particles. For more details, see Ref.~\cite{belle-khh3}.
The $f_0(980)$ lineshape is parametrized with a Flatt\'e
function~\cite{Flatte} with parameters fixed at the values determined in the
analysis of the $\bckpp$ decay~\cite{belle-dcpv}:
$M=0.950\pm0.009(stat.)$~GeV/$c^2$ and coupling constants
$g_{\pi\pi}=0.23\pm0.05(stat.)$ and $g_{KK}=0.73\pm0.30(stat.)$~\cite{note-2}.
An additional amplitude
$f_X(1300)$ is introduced to account for an excess of signal events
observed at $M(\pipi)\simeq1.3$~GeV/$c^2$. As found in Ref.~\cite{belle-khh3},
if approximated by a single resonant state, it is best described by a scalar
amplitude. We fix the mass and width of the $f_X(1300)$ at values determined
in Ref.~\cite{belle-dcpv}: $M=1.449\pm0.013(stat.)$~GeV/$c^2$ and
$\Gamma=0.126\pm0.025(stat.)$~GeV/$c^2$. From an analysis with a larger data
sample~\cite{belle-dcpv}, a contribution from $B^+\to f_2(1270)K^+$ is also
found. However, in this analysis, we do not find a significant signal for
$B^0\to f_2(1270)K^0$ (see below), so we do not put it in the default model
but include this channel when evaluating model uncertainty.


For the non-resonant amplitude~$\Am_{\rm nr}$ we use an empirical
parametrization
\begin{equation}
{\cal A}_{\rm nr}(\knpp) =
      a^{\rm nr}_1e^{-\alpha{s_{13}}}e^{i\delta^{\rm nr}_1} 
    + a^{\rm nr}_2e^{-\alpha{s_{23}}}e^{i\delta^{\rm nr}_2},
  \label{eq:kpp-non-res}
\end{equation}
where $a^{\rm nr}_i$, $\delta^{\rm nr}_i$ and $\alpha$ are fit parameters.
It is worth noting here that a similar parametrization was used not only in
the analysis of $\bckpp$~\cite{belle-khh3} but also in $\bckkk$
decays~\cite{belle-khh3,babar-kkk}. Finally, note that, since in this analysis
we do not distinguish between $B$ and $\bar{B}$ decays, the signal density
function is a incoherent sum
\begin{equation}
   S(\ks\pi^\pm\pi^\mp) = |{\cal{M}}(K^0\pi^+\pi^-)|^2 +
                          |{\cal{M}}(\bar{K}^0\pi^-\pi^+)|^2.
\end{equation}


\begin{figure}[t]
  \centering
  \includegraphics[width=0.48\textwidth]{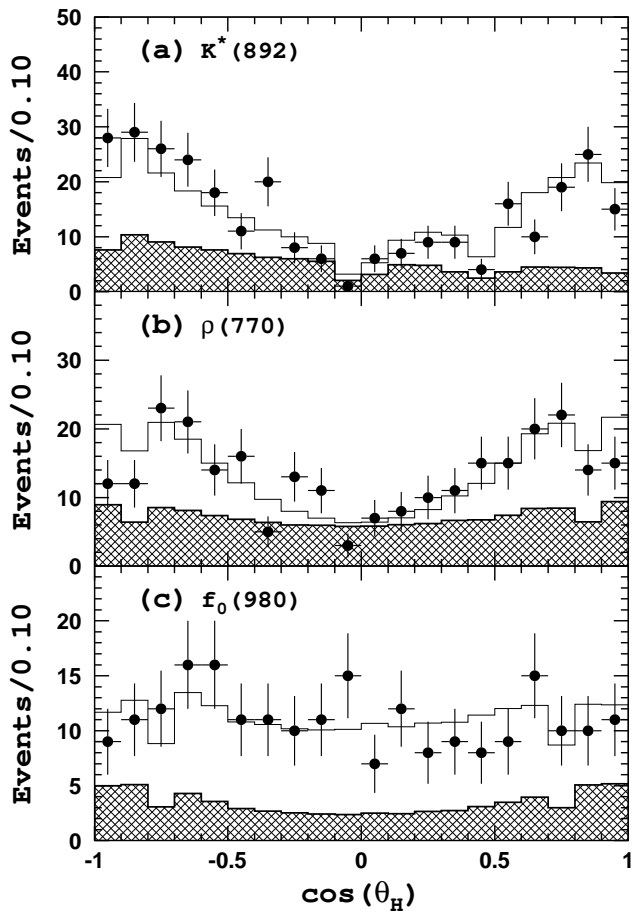}
  \caption{Helicity angle distributions for
           (a)~$K^*(892)$  (0.82~GeV/$c^2$~$<M(\ks\pi)<$~0.97~GeV/$c^2$);
           (b)~$\rho(770)$ (0.60~GeV/$c^2$~$<M(\pi\pi)<$~0.90~GeV/$c^2$);
           (c)~$f_0(980)$  (0.90~GeV/$c^2$~$<M(\pi\pi)<$~1.06~GeV/$c^2$).
           Points with error bars are data, the open histogram is the fit
           result and the hatched histogram is the background component.
           Note that there are two entries per $B$
           candidate in plot (a).}
\label{fig:kpp-heli}
\end{figure}


\begin{table*}[!th]
  \caption{Summary of fit results. The first quoted error is statistical,
           the second is systematic and the third is the model error.
           $R$ in the two last columns denotes an intermediate resonant
           state and $h$ stands for a final state hadron: pion or neutral
           kaon. The $\knpp$ charmless fraction excludes the $B^0\to\chic K^0$
           contribution.}
  \medskip
  \label{tab:results}
\centering
  \begin{tabular}{lcccr} \hline \hline
 Mode &
 $f_i$ \% & $\delta_i$ $^\circ$ &
 $\BF(B\to Rh)\times\BF(R\to hh)\times10^{6}$ &
 $\BF(B\to Rh)\times10^{6}$ 
\\ \hline 
 $\knpp$  charmless~~   &  $99.3\pm0.4\pm0.1$ & $-$ & $-$
                        & $47.5\pm2.4\pm3.7$  \\
 ~~$K^*(892)^+\pi^-$
                        & $11.8\pm1.4\pm0.5^{+0.9}_{-0.6}$  
                        & $0$ (fixed)   
                        & $5.6\pm0.7\pm0.5^{+0.4}_{-0.3}$
                        & $8.4\pm1.1\pm0.8^{+0.6}_{-0.4}$  
\\
 ~~$K^*_0(1430)^+\pi^-$
                        & $64.8\pm3.9\pm0.5^{+1.6}_{-6.3}$
                        & $45\pm9\pm2^{+9}_{-13}$ 
                        & $30.8\pm2.4\pm2.4^{+0.8}_{-3.0}$
                        & $49.7\pm3.8\pm6.7^{+1.2}_{-4.8}$ 
\\
 ~~$K^*(1410)^+\pi^-$
                        & $-$ & $-$
                        & $<3.8$ & \multicolumn{1}{c}{$-$}         \\
 ~~$K^*(1680)^+\pi^-$
                        & $-$ & $-$
                        & $<2.6$ & \multicolumn{1}{c}{$-$}         \\
 ~~$K^*_2(1430)^+\pi^-$
                        & $-$ & $-$
                        & $<2.1$ & \multicolumn{1}{c}{$-$}         \\
 ~~$\rho(770)^0K^0$
                        & $12.9\pm1.9\pm0.3^{+2.1}_{-2.2}$
                        & $-7\pm28\pm7^{+27}_{-13}$ 
                        & $6.1\pm1.0\pm0.5^{+1.0}_{-1.1}$
                        & $6.1\pm1.0\pm0.5^{+1.0}_{-1.1}$ 
\\
 ~~$f_0(980)K^0$
                        & $16.0\pm3.4\pm0.8^{+1.0}_{-1.4}$
                        & $36\pm34\pm5^{+38}_{-21}$ 
                        & $7.6\pm1.7\pm0.7^{+0.5}_{-0.7}$
                        & \multicolumn{1}{c}{$-$}                 
\\
 ~~$f_X(1300)K^0$
                        & $3.7\pm2.2\pm0.3^{+0.5}_{-0.5}$
                        & $-135\pm25\pm2^{+26}_{-31}$ 
                        & $-$
                        & \multicolumn{1}{c}{$-$}                 
\\
 ~~$f_2(1270)K^0$
                        & $-$ & $-$
                        & $<1.4$ & \multicolumn{1}{c}{$-$}         
\\
 ~~Non-resonant
                        & $41.9\pm5.1\pm0.6^{+1.4}_{-2.5}$
                        & $\delta^{\rm nr}_1=-22\pm8\pm1^{+6}_{-6}$
                        & $-$
                        & $19.9\pm2.5\pm1.6^{+0.7}_{-1.2}$  
\\
                        &
                        & $~\delta^{\rm nr}_2=175\pm30\pm4^{+54}_{-30}$
                        &   &  
\\
\hline
 $\chic K^0$
                        & $-$
                        & $-$
                        & $<0.56$
                        & $<113$                                   \\
\hline \hline
  \end{tabular}
\end{table*}

When fitting the data, we choose the $K^*(892)^+\pi^-$ signal as our
reference by fixing its amplitude and phase ($a_{K^*}\equiv 1$ and
$\delta_{K^*}\equiv 0$). Two-particle mass projections for the fit and
data are compared in Fig.~\ref{fig:kpp-sig-fit}. In addition, 
Fig.~\ref{fig:kpp-heli} shows the helicity angle distributions for several
regions. The helicity angle for the $\pi^+\pi^-$ system is defined as the
angle between the $\pi^-$ flight direction and the $B$ flight direction
in the $\pi^+\pi^-$ rest frame. For the $\ks\pi$ system, the helicity
angle is defined with respect to the $\ks$.
All plots in Figs.~\ref{fig:kpp-sig-fit} and~\ref{fig:kpp-heli} demonstrate
good agreement between the fit and data. Results of the fit are summarized in
Table~\ref{tab:results}, where the relative fraction $f_i$ of a 
quasi-two-body channel in the three-body signal is calculated as
\begin{equation}
 f_i = \frac{\int |a_i{\cal A}^i|^2\, d\sft d\sst}
            {\int |{\cal M}|^2\, d\sft d\sst}.
\label{eq:fraction}
\end{equation}
While the relative fraction for a particular quasi-two-body channel depends
only on the corresponding amplitude in the matrix element in
Eq.~\ref{eq:kpp-model}, its statistical error depends on the statistical
errors of all amplitudes and phases. To determine the statistical errors for
quasi-two-body channels, we use a MC pseudo-experiment technique as described
in Ref.~\cite{belle-khh3}.

We find that a significant fraction of the $\bnkpp$ signal is due to a
non-resonant-like decay and is dominated by the $K-\pi$ component of the
non-resonant amplitude in Eq.~(\ref{eq:kpp-non-res}):
$a^{\rm nr}_2/a^{\rm nr}_1=0.20\pm0.11({\it stat.})$.
This is in agreement with the analysis of $\bckpp$. 
The value of the parameter $\alpha=0.154\pm0.033(stat.)$ of the non-resonant
amplitude obtained from the fit also agrees with that determined in the
analysis of charged $B$ meson decay:
$\alpha(\kcpp) = 0.195\pm0.018(stat.)$~\cite{belle-dcpv}.

To determine the reconstruction efficiency for the three-body $\bnkpp$ decay,
we use MC simulation, where events are distributed over phase space according
to the matrix element obtained from the best fit to data. The corresponding
reconstruction efficiency is $(6.71\pm0.03)$\% (including the
$K^0\to\pi^+\pi^-$ branching fraction). Branching fraction results are given
in Table~\ref{tab:results}. Since the nature of the $f_X(1300)$ is not
well understood, and it might in fact be a mixture of several states (for
example, $f_0(1370)$ and $f_0(1500)$), only a relative fraction and phase are
given for the $f_X(1300)K^0$ channel.
Note that the $\bskpp$ signal yield determined from the fit to the $\de$
distribution includes some contribution from the $B^0\to\chic\ks$ decay, which
is not a charmless decay. To correct for this contribution, we multiply the
signal yield by a factor $0.993$ (see Table~\ref{tab:results}) when calculating
branching fractions.

For the final states where no statistically
significant signal is observed, we calculate 90\% confidence level upper
limits $f_{90}$ for their fractions via
\begin{equation}
0.90=\frac{\int_{0}^{f_{90}}G(f,\sigma_f;x)dx}
          {\int_{0}^{\infty}G(f,\sigma_f;x)dx},
\end{equation}
where $G(f,\sigma_f;x)$ is a Gaussian function with the measured mean value
$f$ for a quasi-two-body signal fraction and its statistical error $\sigma_f$.
To account for the model uncertainty we determine the relative fractions with
various parametrizations of the $B$ decay amplitude (see below) and use the
largest value to evaluate the upper limit. To account for the systematic
uncertainty, we decrease the reconstruction efficiency by one standard
deviation.

To assess how well any given fit represents the data, the Dalitz plot is 
subdivided into non-equal bins requiring that the number of events in each
bin exceeds 25. A goodness-of-fit statistic for the multinomial distribution
is then calculated as
$\chi^2=-2\sum^{N_{\rm bins}}_{i=1}n_i\ln\left(\frac{p_i}{n_i}\right),$ where
$n_i$ is the number of events observed in the $i$-th bin, and $p_i$ is the
number of events predicted from the fit~\cite{baker}. The distribution of this
statistic is bounded by a $\chi^2$ distribution with $(N_{\rm bin}-1)$ degrees
of freedom, and one with $(N_{\rm bin}-k-1)$ degrees of freedom,
where $k$ is the number of fit parameters~\cite{chernoff}. 
The $\chi^2/N_{\rm bins}$ value for the best fit is $124.3/112$ with $k=16$
fit parameters. This corresponds to a confidence level between 2\% and 18\%.
The $\chi^2/N_{\rm bins}$ value of the fit to sideband events
is $241.7/197$ with $k=10$ fit parameters. 

To estimate the model uncertainty we modify the matrix element
Eq.~(\ref{eq:kpp-model}) to include an additional quasi-two-body amplitude:
either $K^*(1410)^+\pi^-$, $K^*(1680)^+\pi^-$, $K^*_2(1430)^+\pi^-$ or
$f_2(1270)K^0$ and repeat the fit to data. For none of these channels is a
statistically significant signal found. We also try to fit the data assuming
$f_X(1300)$ is a vector (tensor) state. In this case its mass and width are
fixed at world average values of $\rho(1450)$ ($f_2(1270)$)~\cite{PDG}.
Finally we try several alternative parametrizations of the non-resonant
amplitude $\Am_{\rm nr}$ to estimate the related uncertainty:
\begin{itemize}
 \item{ $
          a^{\rm nr}_1e^{-\alpha \sft }e^{i\delta^{\rm nr}_1}$}
 \item{ $
          a^{\rm nr}_1e^{-\alpha \sft }e^{i\delta^{\rm nr}_1}+
          a^{\rm nr}_2e^{-\alpha \sst}e^{i\delta^{\rm nr}_2}+
          a^{\rm nr}_3e^{-\alpha \sfs}e^{i\delta^{\rm nr}_3}$}
 \item{ $
         \frac{a_1^{\rm nr}}{\sft^\alpha}e^{i\delta_1^{\rm nr}}+
         \frac{a_2^{\rm nr}}{\sst^\alpha}e^{i\delta_2^{\rm nr}}$}
 \item{ $ \frac{\sqrt{s_{13}}}{p_s\cot\delta_B-ip_s}
 + e^{2i\delta_B}\frac{M_{K^*_0}\Gamma_{K^*_0}\frac{M_{K^*_0}}{p_0}}
   {M^2_{K^*_0}-s_{13}-iM_{K^*_0}\Gamma_{K^*_0}\frac{M_{K^*_0}}{p_0}\frac{p_s}{\sqrt{\sft}}};\\
   \cot\delta_B = \frac{1}{ap_s} + \frac{1}{2}rp_s$}
\end{itemize}
The latter parametrization, where $p_0$ ($p_s$) is the momentum of either
daughter particle in the $K^*_0(1430)$ rest frame calculated at the nominal
(current) mass value, and $a$ and $r$ are parameters, is suggested by the BaBar
Collaboration~\cite{babar-kpp}. It is based on results of the partial wave
analysis of elastic $K$-$\pi$ scattering by the LASS collaboration~\cite{LASS}.
In this parametrization the relative fraction and phase between the
$K^*_0(1430)$ amplitude and an underlying broad scalar amplitude (that in the
LASS analysis is referred to as an effective range term and in our analysis
is described by the independent amplitude $\Am_{\rm nr}$) are fixed from LASS
data. However, the use of LASS data is limited to the elastic region (i.e.
below the $K\eta$ production threshold), thus in BaBar's analysis the effective
range term is truncated slightly above the elastic limit and an additional
non-resonant (phase-space) term is introduced to describe an excess of signal
events at higher $M(K\pi)$. In our analysis additional degrees of freedom
introduced by an independent amplitude $\Am_{\rm nr}$ lead to
a second solution with a slightly worse likelihood value but with a much
smaller $K^*_0(1430)\pi$ signal fraction. MC studies
confirm that the presence of the second solution is due to an interplay between
the two $S$-wave components: the $K^*_0(1430)\pi$ and $\Am_{\rm nr}$, and is
not related to the limited experimental statistics. A similar ambiguity was
found in the analysis of $\bckpp$ and $\bckkk$
decays~\cite{belle-khh3,babar-kkk}. However, comparison of the phase shift of
the total $K$-$\pi$ $S$-wave amplitude (which is a coherent sum of 
$K^*_0(1430)$ and $\Am_{\rm nr}$) as a function of $M(K\pi)$ with that
measured by LASS in the elastic region favors the solution with a large
$K^*_0(1430)\pi$ fraction. This is also in agreement with some phenomenological
estimates~\cite{chernyak}.

\begin{table}[t]
\centering
\caption{Contributions to the systematic uncertainty (in percent) for the
         three-body $\bnkpp$ branching fraction.}
\medskip
\label{khh_syst}
  \begin{tabular}{lc}  \hline \hline
  Source~\hspace*{52mm} &  \multicolumn{1}{c}{Error \%}          \\
\hline 
 Efficiency nonuniformity     &     $2.4$         \\
 Event shape requirements     &     $2.5$         \\
 Signal yield  extraction     &     $5.4$         \\
 Charged track reconstruction &     $2.0$         \\
 Particle identification      &     $2.0$         \\
 $\ks$ reconstruction         &     $3.0$         \\
 $N_{\bbbar}$ estimation      &     $1.0$         \\
\hline
 Total                        &     $7.8$         \\
\hline \hline
  \end{tabular}
\end{table}

The dominant sources of systematic error in the determination of the three-body
$\bnkpp$ branching fractions are listed in Table~\ref{khh_syst}. Because of
the non-uniformity of the reconstruction efficiency over the Dalitz plot, the
reconstruction efficiency for the three-body $\bnkpp$ decay determined from
MC is sensitive to the model used to generate signal events. The associated
systematic uncertainty is estimated by varying the relative phases and
amplitudes of the quasi-two-body states within their errors. The systematic
uncertainty due to requirements on event shape variables is estimated from a
comparison of their distributions for signal MC events and $B\to D\pi$ and
$B\to J/\psi K$ events in the data. We estimate the uncertainty in the signal
yield extraction from the fit to the $\de$ distribution by varying the
parameters of the fitting function within their errors. This includes variation
of parameters of the signal function, normalization of the $B\bar{B}$ related
background and the slope and normalization of the $\qqbar$ background function
within their errors. The uncertainty from the particle identification
efficiency is estimated using pure samples of kaons and pions from
$D^0\to K^-\pi^+$ decays, where the $D^0$ flavor is tagged using
$D^{*+}\to D^0\pi^+$. The systematic uncertainty in charged track
reconstruction is estimated using partially reconstructed $D^*\to D\pi$
events and from comparison of the ratio of $\eta\to\pi^+\pi^-\pi^0$ to
$\eta\to\gamma\gamma$ events in data and MC. For the quasi-two-body channels,
additional sources are the uncertainty in parametrization of the distribution
of background events over the Dalitz plot that is estimated by varying the
parameters of the fitting function Eq.~(\ref{eq:kpp_back}) within their errors.
Finally, there is an 11\% uncertainty in the branching fraction for the
$K^*_0(1430)\to K\pi$ decay~\cite{PDG}.

In summary, an amplitude analysis of the three-body charmless $\bnkpp$ decay
is performed for the first time. The results are summarized in
Table~\ref{tab:results}. The analysis reveals the presence of the
$K^*(892)^+\pi^-$, $K^*_0(1430)^+\pi^-$, $\rho(770)^0K^0$ and $f_0(980)K^0$
quasi-two-body intermediate channels for which we measure the branching
fractions. The $B^0\to\rho(770)^0K^0$ branching is measured for the first time.
We also find that a significant fraction of the $\bnkpp$ signal is due to a
non-resonant component; this is consistent with results from the Dalitz
analysis of $\bckpp$ decays. We obtain upper limits on branching fractions for
several other possible channels; these constraints include the first limits
obtained for $K^*(1410)^+\pi^-$, $K^*(1680)^+\pi^-$, $K^*_2(1430)^+\pi^-$ and
$f_2(1270)K^0$.

We thank the KEKB group for the excellent operation of the
accelerator, the KEK cryogenics group for the efficient
operation of the solenoid, and the KEK computer group and
the National Institute of Informatics for valuable computing
and Super-SINET network support. We acknowledge support from
the Ministry of Education, Culture, Sports, Science, and
Technology of Japan and the Japan Society for the Promotion
of Science; the Australian Research Council and the
Australian Department of Education, Science and Training;
the National Science Foundation of China and the Knowledge
Innovation Program of the Chinese Academy of Sciences under
contract No.~10575109 and IHEP-U-503; the Department of
Science and Technology of India; 
the BK21 program of the Ministry of Education of Korea, 
the CHEP SRC program and Basic Research program 
(grant No.~R01-2005-000-10089-0) of the Korea Science and
Engineering Foundation, and the Pure Basic Research Group 
program of the Korea Research Foundation; 
the Polish State Committee for Scientific Research; 
the Ministry of Science and Technology of the Russian
Federation; the Slovenian Research Agency;  the Swiss
National Science Foundation; the National Science Council
and the Ministry of Education of Taiwan; and the U.S.\
Department of Energy.

\newpage

\clearpage
\newpage


\begin{thebibliography}{99} 

\bibitem{hhh-theory}
M.Gronau, J.L.Rosner, \prd{72}{2005}{094031};
S.Fajfer, R.J.Oakes, and T.N.Pham, \plb{539}{2002}{67};
N.G.Deshpande, N.Sinha and R.Sinha, \prl{90}{2003}{061802};
M.Gronau, \prl{91}{2003}{139101};
T.Gershon and M.Hazumi, \plb{596}{2004}{163}.
M.Gronau, D.Pirjol, A.Soni and J.Zupan, hep-ph/0608243,
Submitted to Phys. Rev. D.
\bibitem{belle-dcpv}
 Belle Collaboration, A.Garmash  {\it et al.}, \prl{96}{2006}{251803}.
\bibitem{b2s-belle}
 Belle Collaboration, K-F.Chen {\it et al.}, \prd{72}{2005}{012004}.
\bibitem{kskckc-babar}
 BaBar Collaboration, B.Aubert   {\it et al.}, \prd{71}{2005}{091102}.
\bibitem{ksksks-belle}
 Belle Collaboration, K.Sumisawa {\it et al.}, \prl{95}{2005}{061801}.
\bibitem{ksksks-babar}
 BaBar Collaboration, B.Aubert   {\it et al.}, \prl{97}{2005}{011801}.
\bibitem{f980-babar}
 BaBar Collaboration, B.Aubert   {\it et al.}, \prl{94}{2005}{041802}.
\bibitem{note-1}
 For the most recent update of the results on studies of $CP$ violation
in charmless $B$ meson decays see Heavy Flavor Averaging Group web page: \\
{\tt http://www.slac.stanford.edu/xorg/hfag/index.html}.
\bibitem{KEKB}
 S.Kurokawa and E.Kikutani, \nima{499}{2003}{1}.
\bibitem{Belle}
 A.Abashian {\it et al.}, \nima{479}{2002}{117}.
\bibitem{GEANT}
 R.Brun {\it et al.}, GEANT 3.21, CERN Report DD/EE/84-1, 1984.
\bibitem{belle-khh2}
 Belle Collaboration, A.Garmash {\it et al.}, \prd{69}{2004}{012001}.
\bibitem{belle-khh3}
 Belle Collaboration, A.Garmash {\it et al.}, \prd{71}{2005}{092003}.
\bibitem{PDG}
 Particle Data Group, W.-M.Yao {\it et al.}, \jpg{33}{2006}{1}.
\bibitem{ffactor} M.Wirbel, B.Stech, M.Bauer, \zpc{29}{1985}{637}.
\bibitem{blatt-weisskopf}
 J.Blatt and V.Weisskopf, {\it Theoretical Nuclear 
 Physics}. New York: John Wiley \& Sons (1952).
\bibitem{Flatte}
S.M.Flatt\'e, \plb{63}{1976}{224}.
\bibitem{note-2}
These values of the $f_0(980)$ parameters are consistent with those reported
by the BES collaboration:\\
BES Collaboration, M. Ablikim {\it et al.}, \plb{607}{2005}{243}.
\bibitem{babar-kkk}
 BaBar Collaboration, B.Aubert {\it et al.}, \prd{74}{2006}{032003}.
\bibitem{baker}
 S.Barker and R.Cousins, \nim{221}{1984}{437}.
\bibitem{chernoff}
 H.Chernoff and E.L.Lehman, Ann.\, Math.\, Stat.\, 25, 579 (1954).
\bibitem{babar-kpp}
 BaBar Collaboration, B.Aubert {\it et al.}, hep-ex/0408073.
\bibitem{LASS}
 LASS Collaboration, D.Aston {\it et al.}, \npb{296}{1988}{493}.
\bibitem{chernyak}
 V.L.Chernyak, \plb{509}{2001}{273}.
\bibitem{belle-kpp0}
 Belle Collaboration, P.Chang {\it et al.}, \plb{599}{2004}{148}.

\end{thebibliography}
\end{document}